\documentclass{aa}

\usepackage{graphicx}
\usepackage{txfonts}
\usepackage{natbib}
\usepackage{url}
\begin{document}

   \title{A new procedure for defining a homogenous line-list for solar-type stars\thanks{Table \ref{tab3} is fully available in electronic form
at the CDS via anonymous ftp to cdarc.u-strasbg.fr (130.79.128.5) or via http://cdsweb.u-strasbg.fr/cgi-bin/qcat?J/A+A/}}


   \author{S. G. Sousa\inst{1,}\inst{2,}\inst{4}
          \and
	  N. C. Santos\inst{1,}\inst{2}
	  \and V. Adibekyan\inst{1}
	  \and E. Delgado-Mena\inst{1}
	  \and H.M. Tabernero\inst{3}
	  \and J.I. Gonz\'alez Hern\'andez\inst{4,}\inst{5}
	  \and D. Montes\inst{3}
	  \and R. Smiljanic\inst{6,}\inst{7}
	  \and A. J. Korn\inst{8}
	  \and M. Bergemann\inst{9}
	  \and C. Soubiran\inst{10}
	  \and S. Mikolaitis\inst{11,}\inst{12}
          }

	  \institute{Centro de Astrof\'isica, Universidade do Porto, Rua das Estrelas, 4150-762 Porto, Portugal
	  \and Departamento de F\'isica e Astronomia, Faculdade de Ci\^encias, Universidade do Porto, Rua do Campo Alegre, 4169-007 Porto, Portugal
	  \and Dpto. de Astrofisica, Facultad de CC. Fisicas, Universidad Complutense de Madrid, E-28040 Madrid, Spain
	  \and Instituto de Astrofisica de Canarias, Cl. Via Lactea s/n, E-38200 La Laguna, Tenerife, Spain
	  \and Dept. Astrofisica, Universidad de La Laguna (ULL), E-38206 La Laguna, Tenerife, Spain
	  \and European Southern Observatory, Karl-Schwarzschild-Str. 2, 85748 Garching bei M\"unchen, Germany
	  \and Department for Astrophysics, Nicolaus Copernicus Astronomical Center, ul. Rabia\'nska 8, 87-100 Toru\'n, Poland
	  \and Department of Physics and Astronomy, Division of Astronomy and Space Physics, Uppsala University, Box 516, 75120 Uppsala, Sweden
	  \and Max Planck Institute for Astrophysics, Karl-Schwarzschild str. 1, 85741 Garching
	  \and Université de Bordeaux 1 – CNRS – Laboratoire d’Astrophysique de Bordeaux, UMR 5804, BP 89, 33271 Floirac Cedex, France
	  \and Laboratoire Lagrange (UMR 7293), Université de Nice Sophia Antipolis, CNRS, Observatoire de la Côte d’Azur, BP 4229, 06304 Nice Cedex 04, France
	  \and Institute of Theoretical Physics and Astronomy, Vilnius University, Goštauto 12, Vilnius 01108, Lithuania
}

   \date{}

 
   \abstract
    {The homogenization of the stellar parameters is an important goal for large observational spectroscopic 
    surveys, but it is very difficult to achieve it because of the diversity of the spectroscopic analysis methods 
    used within a survey, such as spectrum synthesis and the equivalent width method. To solve this problem, constraints 
    to the spectroscopic analysis can be set, such as the use of a common line-list.}
    {We present a procedure for selecting the best spectral lines from a given input line-list, which then allows us 
    to derive accurate stellar parameters with the equivalent width method.}
%
    {To select the lines, we used four very well known benchmark stars, 
    for which we have high-quality spectra. From an initial line-list, the equivalent width of each individual line 
    was automatically measured for each benchmark star using ARES, then we performed a local thermodynamic equilibrium analysis with MOOG to 
    compute individual abundances. The results allowed us to choose the best lines which give consistent 
    abundance values for all the benchmark stars from which we then created a final line-list.}
%
%
     {To verify its consistency, the compiled final line-list was tested for 
     a small sample of stars. These stars were selected to cover different ranges in the parameter 
     space for FGK stars. We show that the obtained parameters agree well with previously determined values.}
%
    {}


   \maketitle
%

\section{Introduction}

The detailed characterization of stars is of extreme importance in several fields 
of astrophysics, for instance from the detailed study of individual objects to the wider 
study of many stars, as the one that is needed to identify and understand the different 
populations of stars that are part of our Galaxy's structure. Spectroscopy is a very 
powerful technique for this purpose, and is often explored in many observational surveys.



Large observational surveys typically need for significant and expert man-power to obtain and analyse the data. This extends from the 
preparation of the observations, and the actual collection of the data, to the data reduction.
The final product for such large surveys may include radial velocities, rotational velocities, 
but also very importantly, the homogeneous spectroscopic stellar parameters and detailed chemical abundances 
for the observed targets. 


In this work we focus on the Equivalent Width (EW) method spectroscopic analysis, more specifically, on selecting the lines that best allow us to 
derive homogeneous parameters to study solar-type stars in different evolutionary stages that cover different metallicity regimes. The 
procedure presented here can be easily adapted for any other constraints, such as the use of other atmospheric models or of a different preliminary 
list of iron lines, or even another chemical element.

In Section 2, we review our spectroscopic analysis method based on equivalent widths. 
We describe some of the different approaches that can be adopted, especially 
regarding the atomic data of spectroscopic lines, and describe the way to perform a differential 
analysis. In Section 3, we describe homogenization constraints that may be imposed on large spectroscopic 
surveys. In Section 4 we present the benchmark stars used for the work, referring to 
the best adopted parameters that we can choose from other works. In Section 5, we describe the procedure of selecting the 
best lines using the benchmark stars and the respective adopted parameters. Subsequently, in Section 6 we present some results 
derived with the method and the selected lines, and compare with to other results found in 
literature. In Section 7 we conclude with a brief summary.


\section{Equivalent width method - ARES+MOOG}

There are two main approaches that are commonly used to analyse spectroscopic data. One 
is usually referred to as the synthetic method, which compares synthetic spectra with the 
observed spectrum and find the best reproduction. The other, referred to as EW method, 
uses the opposite approach. It begins directly with the observed spectrum, normally 
analysing it line by line, measuring each line's strength, typically 
through the EW, and computing individual abundances. The two approaches have their 
own advantages and disadvantages. For example, one of the advantages of the EW method 
is its higher efficiency for the spectral analysis, which is focused on a well-defined set of lines, 
while the synthetic method typically requires more complicated computations to generate synthetic 
spectra for comparison or, alternatively, a large grid of pre-computed 
synthetic spectra. On the other hand, one of the advantages of the synthetic method is 
that it can be applied to fast-rotator stars, while the analysis of individual lines 
through the EW method becomes impossible for these cases because of the severe line blending.

\subsection{Standard equivalent width method}

A standard technique for deriving stellar atmospheric parameters using the EW method is based on the iron ionization and excitation balance. 
Iron produces many absorption lines for solar-type stars, and therefore it is used frequently for this method. The standard method starts with the 
adoption of a given stellar atmosphere model, characterized by its stellar effective temperature, surface 
gravity, microturbulence, and metallicity (where [Fe/H] is used as a proxy). Then, using a set of \ion{Fe}{i} and \ion{Fe}{ii} lines, 
iterates these parameters until the correlation coefficients between the abundance of \ion{Fe}{i} with the excitation 
potential (sensitive to temperature) and the reduced EW ($\log(\texttt{EW}/\lambda)$, sensitive to micro turbulence) are zero. The abundances 
derived from the \ion{Fe}{ii} lines (sensitive to the surface gravity) are forced to be equal to those obtained from the \ion{Fe}{i} 
lines (less sensitive to surface gravity). The final atmospheric parameters, and the resulting iron abundances may be 
found in this way. 

There are several processes to find the best stellar atmosphere that fits the measured EWs. One simple approach is to find each parameter 
independently, step by step, or, using a more correct approach, to apply complex minimization algorithms that allow one to keep the parameters 
free and that better explores all the unknown interdependences between the parameters. The EW method has been applied in the past 
by several other groups \citep[e.g.][]{Santos-2004b, Sousa-2008, Ghezzi-2010, Mucciarelli-2013, Magrini-2010, Tabernero-2012}


\subsection{Equivalent width measurement with ARES}

The measurement of EWs is one of the main key points. In many previous works, this measurement has been 
performed using interactive routines (e.g. \texttt{splot} on IRAF), where it is necessary to define the position of the 
continuum manually, and tp identify all the absorption lines that are present and are required to be correctly identified for a good and proper 
fitting in the case of blended lines. This procedure, repeated 
individually for each line in each spectrum, is not only very time-consuming, but more importantly, is a subjective procedure, and therefore 
susceptible to provide different systematics on the measurements.
More recently, many authors have started to use automatic tools to measure the EWs (e.g. ARES \citep[][]{Sousa-2007}, DAOSPEC \citep[][]{Stetson-2008}). 
The advantages of these tools include the higher efficiency in obtaining the measurements for many lines, and they allow one 
to keep the individual measurements consistent regarding the continuum determination and the identification of the blended 
lines to be fitted. 
However, there are disadvantages to the automatic tools, namely the uncertainty that we can have in each individual measurement, which 
can be overcome either with a very good selection of the lines to be analyzed, or by using large set of lines that can be used together to identify 
the poor measurements (e.g. using an outlier removal when computing the mean abundance values).

\subsection{Abundances with MOOG}

We used MOOG\footnote{http://www.as.utexas.edu/$\sim$chris/moog.html} \citep{Sneden-1973} to compute the line-by-line abundance 
for each star assuming local termodynamic equilibrium. In our standard method we used a grid of Kurucz Atlas\,9 plane-parallel 
model atmospheres \citep[][]{Kurucz-1993} to 
generate the stellar atmosphere model that was inserted into MOOG to compute the abundances through the driver \textit{abfind}.

\begin{figure}[t]
\centering
  \includegraphics[width=8.8cm]{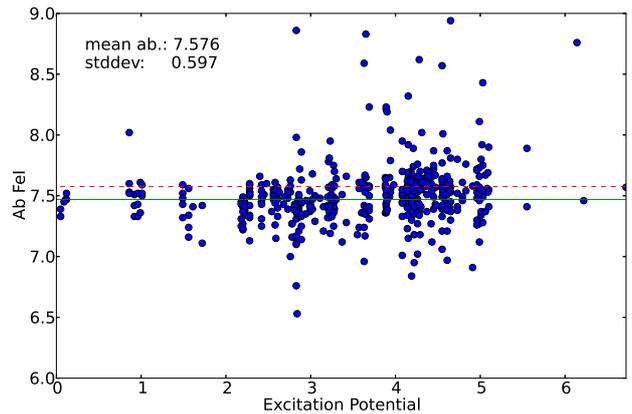}
  \caption{Abundance derived for a selection of iron lines with atomic data taken directly from VALD and EWs derived from a solar spectrum. 
  The full-drawn line represents the iron solar abundance as 7.47. The dashed line represents the mean abundance.}
  \label{Fig1}
\end{figure}

\begin{table*}[ht]
\caption{Reference stars and adopted parameters}             
\label{tab1}      
\centering          
\begin{tabular}{| c c c c c c c | c c c |}     
\hline       
\multicolumn{7}{|c|}{Adopted values} & \multicolumn{3}{|c|}{Stats from PASTEL}\\
ID & Star & Teff & logg & [Fe/H] & Log(Fe) & $\xi_{\mathrm{t}}$ & $<Teff>$ & $<log g>$ & $<[Fe/H]>$ \\ 
   &      &  (K) &  dex &   dex  &    dex     &     km/s           &    (K)   &     dex   &    dex     \\ 
\hline                    
\hline
   MRD & Sun       & 5777   & 4.44 &  0.00  & 7.47  & 1.00 &    -     &     -      &     -       \\
   MPD & $\mu$ Cas & 5320   & 4.45 & -0.83  & 6.64  & 0.90 & 5315(90) & 4.49(0.18) & -0.82(0.12) \\  
   MRG & $\mu$ Leo & 4540   & 2.10 &  0.15  & 7.62  & 1.80 & 4495(97) & 2.38(0.25) &  0.22(0.18) \\
   MPG & Arcturus  & 4247   & 1.59 & -0.54  & 6.93  & 1.80 & 4316(62) & 1.68(0.31) & -0.54(0.10) \\
\hline                  
\end{tabular}
\end{table*}

\subsection{Differential analysis}

Another important step to define this method is the selection of the lines 
for the individual spectral analysis. The number of lines selected for the method 
has strongly increased with the new automatic tools. 
However, some caution needs be taken and one needs to determine whether all the selected lines are 
suitable for an automatic measurement for the different spectral types of stars. A procedure to select ``stable'' 
lines has been used in \citet[][]{Sousa-2008}. Using a sample of different stars with preliminary parameters, 
each abundance derived from each automatically measured line was compared with the average abundance, identifying 
this way all unstable lines which were systematically given higher or lower relative abundances.

The atomic data for each selected line is fundamental for deriving accurate and precise spectroscopic parameters. Typically, one uses 
atomic data derived in the laboratory or the atomic data used before by other authors.
The differential analysis was invented because the atomic parameters available for the lines are not 
 only poorly determined. These values can be derived in the laboratory, 
but can have large uncertainties of the order of 10-20\%, which can produce a 
wide dispersion in the abundance determination. As an example, we 
present in Fig. \ref{Fig1} the abundance derived for a selection of iron lines with 
atomic data taken directly from the Vienna Atomic Line Data-base 
(VALD: \citet[][]{Piskunov-1995, Kupka-1999,Ryabchikova-1999} - 
http://www.astro.uu.se/htbin/vald) and EWs derived from 
a solar spectrum. Although we are quite certain of the solar atmospheric 
parameters and iron abundance for the Sun, whose values varying around 7.47 
\citep[][]{Anders-1989, Lodders-2009,Asplund-2009},
and errors of the order of 0.05 dex, the figure clearly shows a huge 
dispersion ($\sim$ 0.60 dex) and a significant offset for the expected iron abundance 
($\sim$ 7.58). This high dispersion and offset may be partially explained by the significant number of lines that 
give an overestimated abundance in the top-right part of the figure. These lines should of course be discarded or corrected 
for in a proper spectral analysis. This simple exercise clearly shows that a large portion of the adopted atomic data 
values are not adequate for an accurate and precise analysis.

To correct for these uncertainties a 
benchmark star with very well constrained parameters is typically compared with the star to be studied. In our case 
the benchmark star typically was the Sun, and 
therefore it is expected that this procedure works well for solar-type stars.

Assuming the same atomic data for every star, the trick in this analysis is to 
recompute the astrophysical oscillator strength 
values ($\log gf$) using an inverse analysis. We first measured the 
EW for our selected lines. This was normally done using 
very high quality spectra of the benchmark star, ideally
obtained with the same instrument and in the same conditions as used to observe 
the other stars. 
Secondly, assuming stellar parameters for our benchmark 
star (e.g. for the Sun: $T_{eff}$ = 5777K, log g = 4.44 dex, 
$\xi_{\mathrm{t}}$ = 1.00 $kms^{-1}$ , and 
log(Fe) = 7.47), the values for each $\log gf$ were then changed until 
we derived the observed EW value for the line measured before. 
This process can be carried out using MOOG with the \textit{ewfind} driver 
in an iterative way for each line in the list.

Using a spectroscopic differential analysis relative to the Sun, one can eliminate both the errors on the atomic 
parameters and also part of the measurement errors on the 
equivalent widths that are still present. When we measure the lines 
in the benchmark star we also include errors given by 
the spectrum itself. An example are small 
blends, or the inaccurate position of the continuum in that region of the spectrum. 
When we compute the $\log gf$ for these lines we include all these 
errors, which may partially compensate for the errors for the 
systematic measurement of the EW of the same line for the stars that are analysed. 

\subsection{Applications}

The method described here has been applied by our team in several different 
works in the past, allowing us to derive homogeneous 
and precise parameters for studying extrasolar planets 
\citep[][]{Santos-2004a,Sousa-2006,Sousa-2007,Sousa-2008,Sousa-2011a,
Sousa-2011b, GomezMaqueoChew-2013} and deriving homogeneous chemical 
abundances with the same tools \citep[][]{Neves-2009, DelgadoMena-2010, 
GonzHern-2010, Adibekyan-2012}. It is 
also being used for the follow-up of asteroseismic targets, where its 
accuracy has been tested \citep[][]{Dogan-2013,Huber-2012,Grigahcene-2012,
Creevey-2012}. In many of these works, several comparisons with other 
methods have been presented, showing generally consistent results.

\section{Homegenization constraints}

\begin{figure*}
  \centering
  \includegraphics[width=18cm]{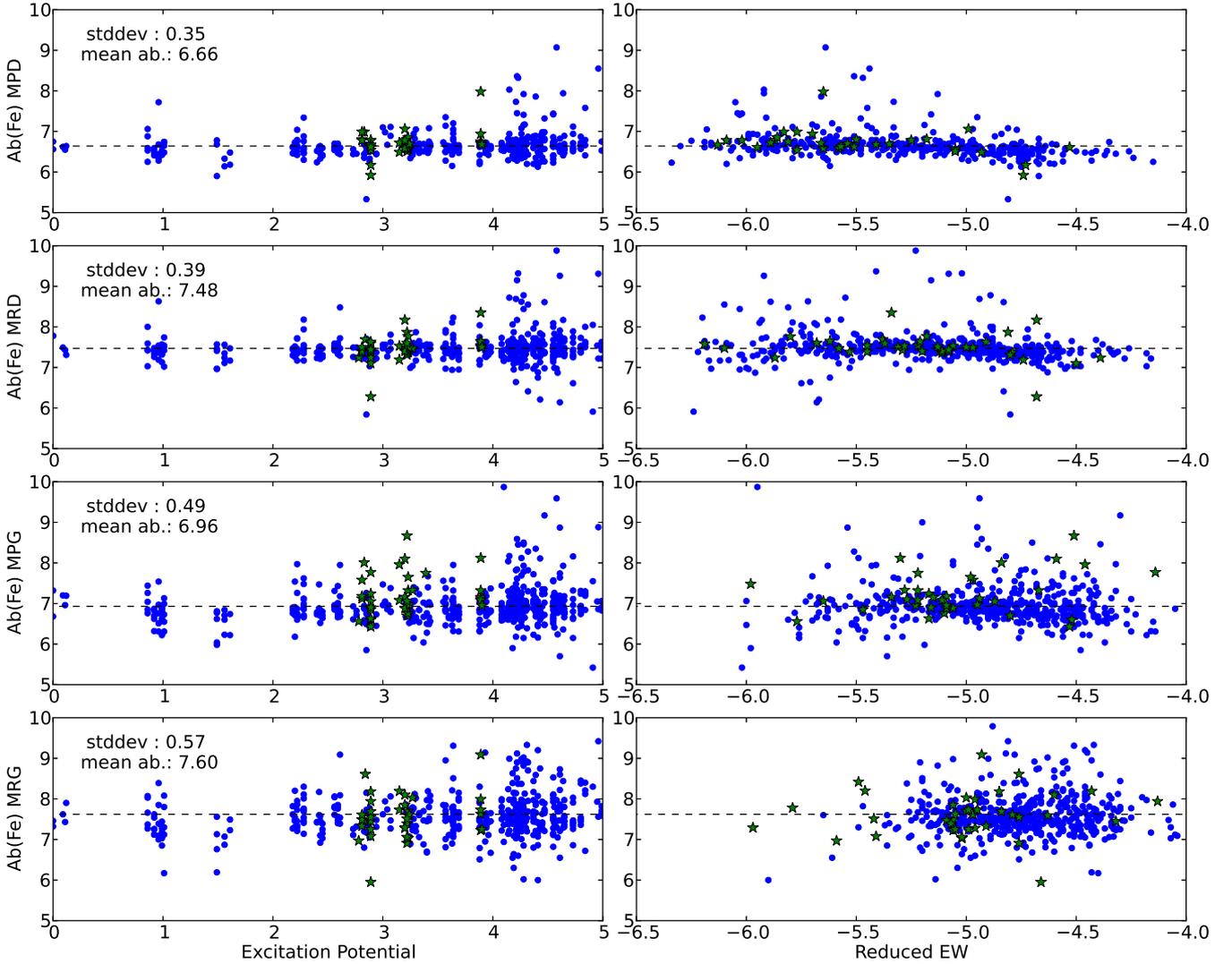}
  \caption{For each benchmark star we present the abundance as a function of excitation potential (left panels) and 
  reduced EW (right panels). Filled circles represent \ion{Fe}{i} lines, while the filled stars represent \ion{Fe}{ii} lines. 
  For each benchmark star the mean iron abundance and the respective standard deviation are indicated.}
  \label{Fig2}
\end{figure*}

\subsection{MARCS models}

For this work we chose to use the MARCS models\footnote{http://marcs.astro.uu.se/} 
from Uppsala \citep[][]{Gustafsson-2008}. 
This is a grid of one-dimensional, hydrostatic, plane-parallel, and spherical 
LTE model atmospheres that may be used together with atomic 
and molecular spectral-line data and software for radiative transfer to generate 
synthetic stellar spectra. Alternatively, the models can be adopted to be used in MOOG, 
which together with derived EWs can be use to compute the respective abundance.

To use MARCS models in our standard automatic procedure, it was necessary 
to adopt the interpolation code available on the MARCS models web page. This 
interpolation code needs to be fed by the correct models around the interpolation 
point in the space of parameters. The selection of the correct models needs to be 
provided as an input. Therefore, to automatize this process, we built 
a script\footnote{The script will be made available by the authors upon request} around 
the provided interpolation code. The simple job of 
this script is to receive the parameters for which one requests a MARCS atmospheric 
model, and perform the correct and automatic selection of models for the interpolation.

There are two different grids for the MARCS models. One assumes plane-parallel 
approximation, the other assuming spherical symmetry. The first is more adequate 
for modelling the atmosphere of dwarf stars, the second is more appropriate for giant stars.
The interpolation code jumps between the grids 
depending on the surface gravity that is asked for the interpolation. Lower surface 
gravities ($ \log g < 3.5$) will use the spherical approximation grid, while for the higher 
log g we used the plane-parallel grid. However, MOOG interprets spherical models as plane parallel, which 
may introduce some differences between the two MARCS grids. These differences are not important for the F, 
G, and K giants \citep[see][]{Heiter-2006}.

Finally, a slight modification was made to the provided interpolation code to 
have the output interpolated model in the format needed for it to be correctly 
applied to MOOG.

\begin{figure*}
  \centering
  \includegraphics[width=18cm]{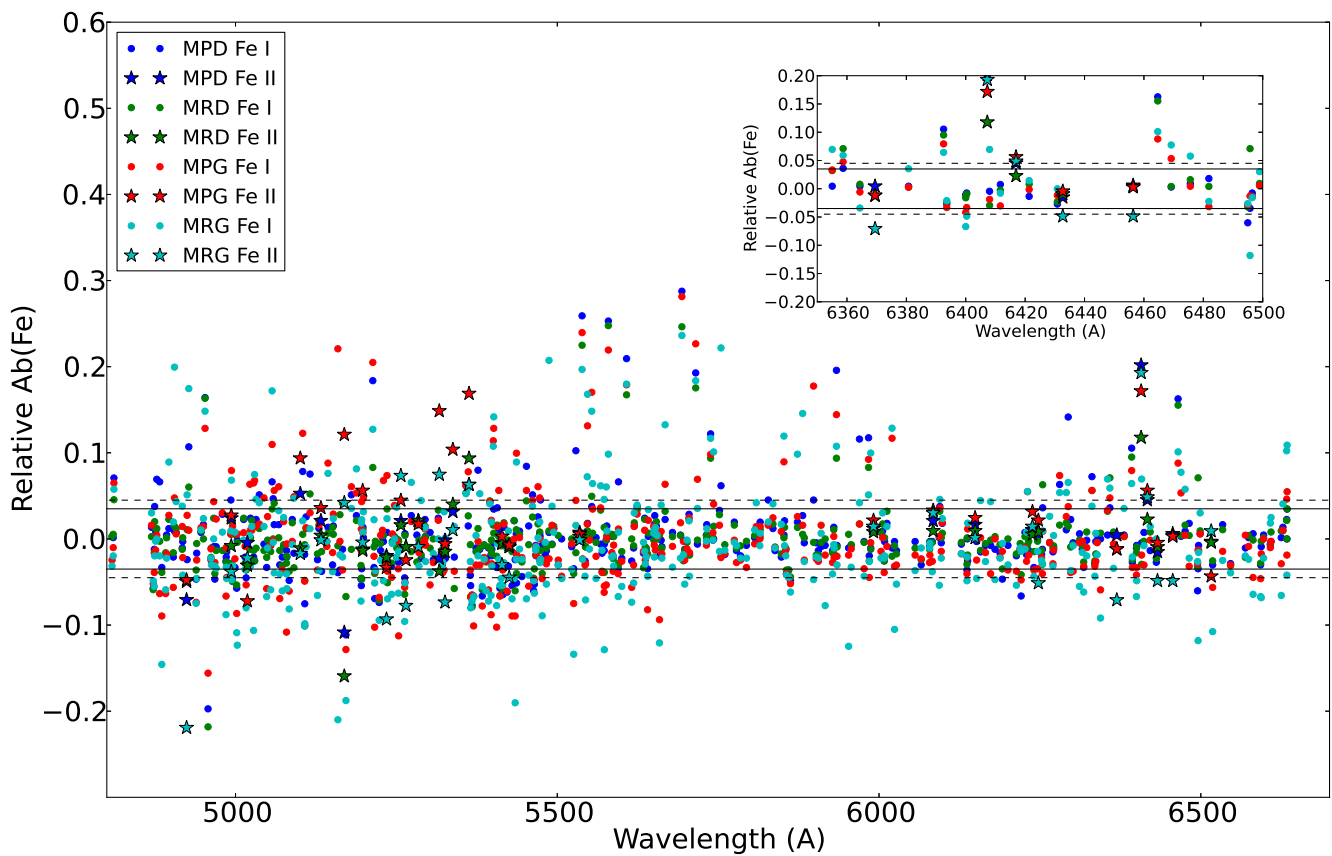}
  \caption{Relative abundance for each iron line for the four benchmark stars. The filled lines and the dashed lines 
  represent the threshold for acceptance of \ion{Fe}{i} and \ion{Fe}{ii} lines, respectively. MPD - metal poor dwarf; MRD - 
  metal rich dwarf; MPG - metal poor giant; MRG - metal rich giant.}
  \label{Fig3}
\end{figure*}

\subsection{Master line-list}

An important constraint that may be imposed for large spectroscopic analysis 
is the use fix atomic data making impossible to use a differencial analysis like in 
our standard method. For these work we use the master linelist defined within 
the Gaia-Eso Survey (GES - \citep[][]{Gilmore-2012}). The input master linelist was kindly 
provided by the GES linelist subworking group prior to publication (Heiter et al. in prep.).

The fixed atomic data adopted for each element line that is used for the whole 
analysis process is of course an important step for the homogenization, but it 
is also a drawback for our standard method where we used a 
differential analysis, as described before. This forced our group to 
use an alternative approach, which is described in detail in the following sections. 

We emphasize here that if correctly adapted, the procedure presented 
here can also be used for other types of stars, which may require different atmospheric models and a 
different set of lines to start with. Therefore the initial line-list may be compiled from whatever 
source and the analysis can also be performed using other model atmospheres (e.g. Kurucz models).


\begin{figure*}
  \centering
  \includegraphics[width=18cm]{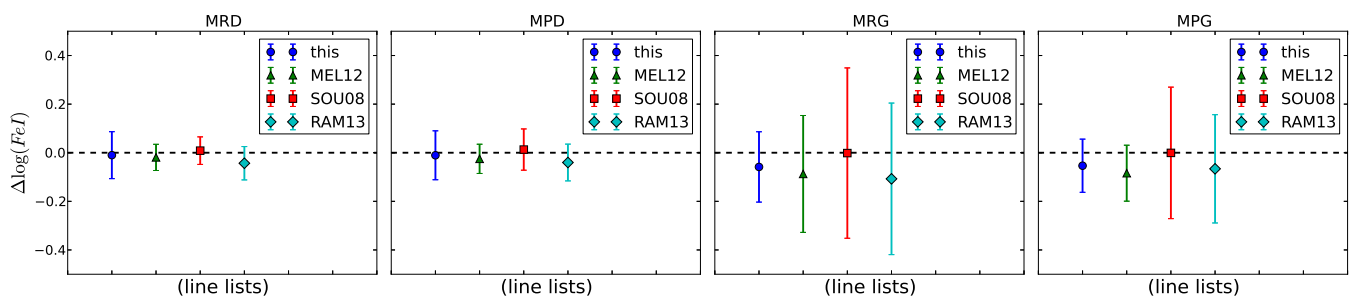}
  \caption{Difference between the mean iron abundance derived for each line-list and the iron abundance
adopted in Table 1 for each star ($\Delta \log(feI)$). The line-lists used for this comparison are presented in \citet[][]{Sousa-2008} (SOU08), 
\citet[][]{Melendez-2012} (MEL12), \citet[][]{Ramirez-2013} (RAM13), and the linelist derived with the procedure described in this work (this).}
  \label{Fignew}
\end{figure*}

\section{Benchmark stars}

The benchmark stars were selected from an exercise developed within a work group for spectra analysis. There are four different 
stars covering different ranges of metallicity and evolutionary stages: MPD - metal-poor dwarf; 
MRD - metal-rich dwarf; MPG metal-poor giant; and MRG metal-rich giant. 
This exercise was made by several nodes within the work group. A detailed 
description of this exercise and respective results will be presented in a separate 
work (Smiljanic et al. in prep.). As stated before, the work 
presented here was developed in the framework of the work group. In this section 
we only intend to present the benchmark stars used to calibrate our procedure.

\subsection{Adopted parameters}

We used the same four stars that were provided for the exercise. They were 
selected in a way to test the different methods in different parameter regimes. 
The four stars are clearly divided into two regimes: a metallicity regime (metal-rich 
vs. metal-poor), and evolutionary-state regime (dwarf vs. giant). This last regime 
strongly depends on temperature, since the giants that are observable are usually cooler stars.

For the Sun we adopted the typical values used by our team in previous works, 
for the other stars we considered the values adopted for the exercise in the work group and 
values from the literature for each star. We used the PASTEL 
catalogue \citep[][]{Soubiran-2010} to find the range of values for the parameters 
of the benchmark stars. Using only 
the rows of the catalogue that present three parameters ($T_{eff}$, logg and [Fe/H]), we show 
the respective standard deviation of the values in Table \ref{tab1}. It is 
clear that the dispersion is quite large, particularly for the surface gravity and the iron 
abundance, which shows how difficult it is to be consistent when using different methods even if 
sometimes the same data are used. From the several values from the literature, 
we selected those that lie close to the average presented in the PASTEL catalogue.
Therefore the adopted values are well within the interval of values from the literature.
An exception is the gravity of $\mu$ Leo, for which there were two rows from the PASTEL catalogue 
with very high values, pushing the average for higher (overestimated) values. Interestingly, these same rows also present higher 
values for the temperature.

\subsection{Spectral data}

The spectra used for the Sun (MRD) and Arcturus (MPG) are the visible and near-infrared atlases at very high 
resolution (R > 100000) of \citet[][]{Hinkle-2000}. The MPD mu Cas was observed with NARVAL, the spectropolarimeter on 
the 2m Telescope Bernard Lyot at the top of Pic du Midi. It provides complete coverage of the optical spectrum (from  370 to 1050 nm) 
in a single exposure with a resolving power of 78000 on average in spectroscopic mode. The MRG mu Leo was observed with ESPaDOnS 
at the 4m Canada-France-Hawaii Telescope, in polarimetric mode at a resolving power of R $\sim$ 68 000. NARVAL and ESPaDOnS are twin 
spectropolarimeters with the same spectral coverage and the same reduction pipeline.
The four spectra have a high signat-to-noise ratio (> 600). The spectra were also prepared to cover the wavelength range provided by UVES. Details can be 
found in Blanco Cuaresma et al. (in prep.) and in Smiljanic et al. (in prep.).

The only necessary treatment performed on the provided spectra was correction for the radial velocity, 
which is essential to identify the lines in the ARES code.

\section{Line selection procedure}

\begin{figure*}
  \centering
  \includegraphics[width=18cm]{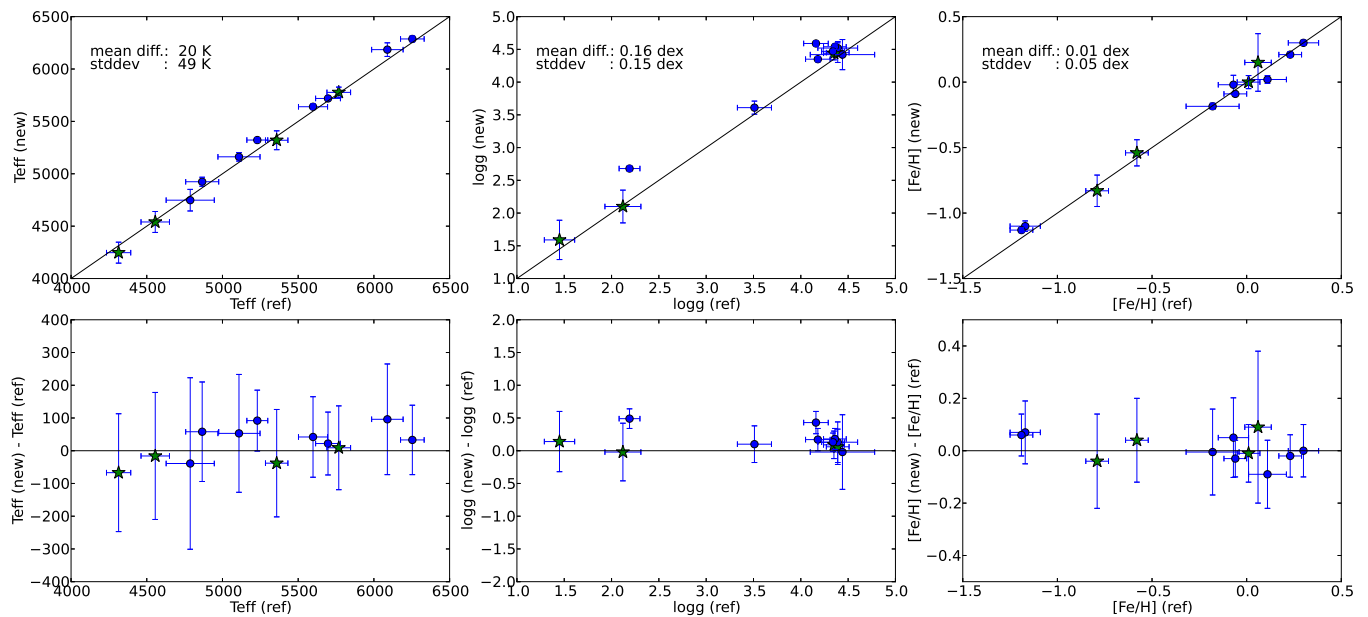}
  \caption{Comparison between the derived (new) and the reference spectroscopic stellar parameters 
  (ref). The star symbols represent the benchmark stars used to select the lines. The filled squares 
  represent the extra stars used to test the ARES+MOOG method with the new line-list.}
  \label{Fig4}
\end{figure*}

As noted before, if one of the constraints is to fix the atomic data, 
we are unable to use our method, which is based on a differential analysis. Since we of course 
use the same fixed atomic data for the lines as everyone else, we therefore  
adopted a special procedure to select the best lines to derive accurate spectroscopic 
parameters. For that purpose, we used the benchmark stars presented before. The procedure 
of finding the best lines is quite simple: adopting fix parameters for the stars, 
we then determine those lines that give reliable iron abundance for all the stars, and 
discard the lines that yield values with a significant offset due to 
either atomic parameters, or poor automatic measurements of the EWs.

\begin{table}[!ht]
\centering 
\caption[]{Sample of the final derived line-list.}
\begin{tabular}{lccc}
\hline
\hline
\noalign{\smallskip}
$\lambda$ (\AA) & $\chi_{l}$ & $\log{gf}$ & Ele.\\
\hline
4808.15 &   3.25 &   -2.690 &   FeI \\
4809.94 &   3.57 &   -2.620 &   FeI \\
4869.46 &   3.55 &   -2.420 &   FeI \\
4875.88 &   3.33 &   -1.920 &   FeI \\
4892.86 &   4.22 &   -1.290 &   FeI \\
4946.39 &   3.37 &   -1.170 &   FeI \\
4950.11 &   3.42 &   -1.670 &   FeI \\
4961.91 &   3.63 &   -2.190 &   FeI \\
4962.57 &   4.18 &   -1.182 &   FeI \\
4969.92 &   4.22 &   -0.710 &   FeI \\
4979.59 &   3.64 &   -2.677 &   FeI \\
4985.55 &   2.87 &   -1.340 &   FeI \\
4999.11 &   4.19 &   -1.640 &   FeI \\
5012.69 &   4.28 &   -1.690 &   FeI \\
5023.19 &   4.28 &   -1.500 &   FeI \\
...     &   ...  &    ...   &   ... \\

\hline
\end{tabular}
\label{tab3}
\end{table}

\subsection{Equivalenth width measurements}

For the four benchmark stars we automatically measured the EWs for all the iron lines included 
in the master line-list that were detected in all the spectra. The 
EWs were measured with the ARES code following the standard input parameters (smoothder=4 - the 
recommended parameter for the smoothing of the derivatives used for the lines identification, 
space=3.0 - the interval in wavelength for each side of the central line that is used for the computation, 
rejt=0.997/0.998 - the important parameter that is used to determine the local continuum which 
strongly depends on the signal-to-noise ratio, lineresol=0.1 \AA- the minimum resolution accepted by ARES to separate 
individual lines). The details of these parameters can be found in \citet[][]{Sousa-2007}. We started with a total of 
586 iron lines and, as expected, the number of detected lines in the benchmark stars strongly 
depends on their metallicity. This effect is compensated for in the case of the metal-poor giants, 
because giant stars in general have stronger lines.

Note that the automatic measurements with ARES are by fitting Gaussian profiles. This 
is acceptable for weak absorption lines (EW $<$ 200 m\AA). At this point we chose to keep 
all the lines, but in our standard procedure we neglected the stronger lines in the spectroscopic analysis 
to avoid systematics caused by poor fitting of these lines. We also neglected very weak lines (EW $<$ 10 m\AA) that 
are strongly affected by sources of error.

\subsection{Line-by line abundances}

Using the adopted parameters for each benchmark star (see Table \ref{tab1}), we interpolated a MARCS model to 
compute the individual iron abundance (using MOOG) for each iron line presented in the 
common line-list. The result is presented in Figure \ref{Fig2}. Although the observed dispersion is 
quite large,  
the mean iron abundances derived for each star are quite consistent with the adopted values. We also 
note that the dispersion is larger for the giant stars. This is because these stars are 
cooler and therefore have a higher density of lines in their spectra, which makes it more difficult to measure individual 
lines because of the stronger blending effects. In addition, the lines in giant stars become stronger, 
and the Gaussian fit that we used for these stronger lines is not as good as that for weaker lines.
More importantly, it is fundamental that no evident correlations are found between the abundances and excitation potential or the reduced 
EW, meaning that the adopted parameters presented in Table \ref{tab1} were chosen well.

\subsection{Selecting the best lines}

Figure \ref{Fig3} shows the relative abundance for each iron line as a function of wavelength 
for each benchmark star. Only the lines that were simultaneously found in the spectra of all 
four benchmark stars are present in this figure, that is, lines that cannot be measured in 
at least one benchmark star were immediately discarded. We then set a value for the 
percentage threshold to accept the best lines. These lines should present a 
relative abundance within 
the threshold for the four benchmark stars simultaneously. After testing with different values, 
we adopted 3.5\% and 4.5\% as the threshold to accept \ion{Fe}{i} lines and \ion{Fe}{ii} lines, respectively. 
The threshold for \ion{Fe}{ii} lines is higher to keep a significant number of these lines 
since they are essential for determining the surface gravity using the ionization equilibrium.

Using this simple procedure, we compiled a final line-list composed of 140 \ion{Fe}{i} lines and 11 \ion{Fe}{ii} 
lines\footnote{The final line-list is available online at CDS, and at \url{http://www.astro.up.pt/~sousasag/sousa2013_table/} }. 
A sample of the line-list is presented in Table \ref{tab3}.
These lines have reliable atomic data, as is shown for the benchmark stars, and are present in the 
different types of stars covered by the benchmark stars used in the process.

\section{Testing the new method}

\begin{table*}[ht]
\caption{Stars used for the test and respective stellar parameters.}             
\label{tab2}      
\centering          
\begin{tabular}{| c | c c c c | c c c c |}     
\hline       
 & \multicolumn{4}{|c|}{Reference Parameters} & \multicolumn{4}{|c|}{New Results}\\
Star & Teff & logg & [Fe/H] & $\xi_{\mathrm{t}}$ & $Teff$ & logg & [Fe/H] & $\xi_{\mathrm{t}}$\\ 
     &  (K) &  dex &   dex  &     km/s           &   (K)  &  dex &   dex  &     km/s          \\ 
\hline                    
\hline
   MRD     & 5777 $\pm$ 50  & 4.44 $\pm$ 0.10 &  0.00 $\pm$ 0.05  & 1.00 $\pm$ 0.05 & 5768 $\pm$ 78  & 4.35 $\pm$ 0.11 &  0.01 $\pm$ 0.06 & 0.90 $\pm$ 0.07 \\
   MPD     & 5320 $\pm$ 90  & 4.45 $\pm$ 0.15 & -0.83 $\pm$ 0.12  & 0.90 $\pm$ 0.20 & 5358 $\pm$ 74  & 4.39 $\pm$ 0.12 & -0.79 $\pm$ 0.06 & 0.65 $\pm$ 0.11 \\
   MRG     & 4540 $\pm$ 100 & 2.10 $\pm$ 0.25 &  0.15 $\pm$ 0.22  & 1.80 $\pm$ 0.20 & 4556 $\pm$ 94  & 2.12 $\pm$ 0.19 &  0.06 $\pm$ 0.07 & 1.84 $\pm$ 0.06 \\
   MPG     & 4247 $\pm$ 100 & 1.59 $\pm$ 0.30 & -0.54 $\pm$ 0.10  & 1.80 $\pm$ 0.20 & 4314 $\pm$ 80  & 1.45 $\pm$ 0.16 & -0.58 $\pm$ 0.06 & 1.75 $\pm$ 0.03 \\
HD125455   & 5162 $\pm$ 41  & 4.52 $\pm$ 0.10 & -0.19 $\pm$ 0.02  & 0.70 $\pm$ 0.10 & 5109 $\pm$ 139 & 4.39 $\pm$ 0.21 & -0.18 $\pm$ 0.14 & 0.22 $\pm$ 1.17 \\
HD179949   & 6287 $\pm$ 28  & 4.54 $\pm$ 0.04 &  0.21 $\pm$ 0.02  & 1.36 $\pm$ 0.03 & 6254 $\pm$ 78  & 4.36 $\pm$ 0.12 &  0.23 $\pm$ 0.12 & 1.13 $\pm$ 0.06 \\
HD2071     & 5719 $\pm$ 14  & 4.47 $\pm$ 0.02 & -0.09 $\pm$ 0.01  & 0.95 $\pm$ 0.01 & 5697 $\pm$ 82  & 4.34 $\pm$ 0.12 & -0.06 $\pm$ 0.06 & 0.61 $\pm$ 0.10 \\
HD72769    & 5640 $\pm$ 27  & 4.35 $\pm$ 0.04 &  0.30 $\pm$ 0.02  & 0.98 $\pm$ 0.03 & 5598 $\pm$ 96  & 4.18 $\pm$ 0.13 &  0.30 $\pm$ 0.08 & 0.94 $\pm$ 0.13 \\
HD102200   & 6185 $\pm$ 65  & 4.59 $\pm$ 0.04 & -1.10 $\pm$ 0.04  & 1.52 $\pm$ 0.23 & 6089 $\pm$ 104 & 4.16 $\pm$ 0.13 & -1.17 $\pm$ 0.08 & 0.59 $\pm$ 0.17 \\
HD221580   & 5322 $\pm$ 24  & 2.68 $\pm$ 0.04 & -1.13 $\pm$ 0.02  & 1.97 $\pm$ 0.04 & 5230 $\pm$ 69  & 2.19 $\pm$ 0.11 & -1.19 $\pm$ 0.06 & 1.43 $\pm$ 0.04 \\
HD22918    & 4924 $\pm$ 43  & 3.61 $\pm$ 0.10 &  0.02 $\pm$ 0.03  & 0.73 $\pm$ 0.08 & 4866 $\pm$ 109 & 3.51 $\pm$ 0.18 &  0.11 $\pm$ 0.10 & 0.36 $\pm$ 0.33 \\
HD105671   & 4748 $\pm$ 103 & 4.42 $\pm$ 0.23 & -0.02 $\pm$ 0.07  & 0.90 $\pm$ 0.34 & 4787 $\pm$ 159 & 4.44 $\pm$ 0.34 & -0.07 $\pm$ 0.08 & 1.18 $\pm$ 0.29 \\
\hline
\end{tabular}
\end{table*}

\subsection{Comparing with other linelists}

To test the line-list derived with the procedure presented here, we compared it 
with other line-lists used in other works to derive of parameters of 
solar-type stars. The line-lists chosen for this comparison are presented in \citet[][]{Sousa-2008} (here as SOU08), 
\citet[][]{Melendez-2012} (here as MEL12), and \citet[][]{Ramirez-2013} (here as RAM13). The test consisted of using 
the benchmark stars to derive the \ion{Fe}{i} abundance for each one of the line-lists, including the 
one that we defined in this work. To computate these abundances we considered that 
the different line-lists were compiled for the use of different models (namely Kurucz and Marcs models) using each 
model for the computation accordingly and assuming the adopted values presented in Table \ref{tab1}. Moreover, the 
MEL12 and SOU08 line-lists adapted their log gf to fit the Sun (differential analysis), while in the 
other line-lists the log gf were adopted from other sources and were fixed (see respective references for more details).

Figure \ref{Fignew} shows the difference between the mean iron abundance derived for each line-list and the 
iron abundance adopted in Table \ref{tab1} for each star. The error bars reflect the dispersion of the abundances 
for each line-list. The values presented in the figure were obtained after a three sigma clipping for each line 
list. Like in the standard methods this allowed us to remove line 
measurements either because of bad continuum fits, or strong blending effects. All 
line-lists yield consistent abundances for the four benchmark stars. As expected for 
the dwarf stars and specifically for the Sun, the dispersion from SOU8 and MEL12 is quite small 
because they used a differential analysis. We also point out that the small difference in abundance seen for 
the SUN in MEL12 is probably related to the different adopted solar iron abundance. For the giant 
benchmark stars the trend is opposite: these two line-lists, together with the RAM13, present the largest 
dispersions of the abundances. From the figure we see that the line-list presented here is appropriate to derive homogeneous 
stellar parameters for these four types of stars. This result comes from the process itself, but it is 
interesting to see the clear increase of the dispersion for the other line-lists from dwarfs to cool giant 
stars.

We would also like to point out that for our line-list the number of lines used for 
the four stars remains nearly constant, while the same does not occur for the other line-lists, where many of 
the lines are not identified (e.g. too weak in metal-poor stars), or there are lines that became so strong in 
giants that they needed to be discarded for a proper analysis. This is an extra additional point in favour of 
the line-list compiled with this process in terms of homogenization for the parameters derived for different stars.

\subsection{Testing with other stars}

The ARES+MOOG method was applied using the MARCS models and the list of lines defined as discussed above. 
To test the method, spectroscopic parameters were derived for the benchmark stars and for eight other stars. 
These stars were selected from previously studied HARPS samples for which high-quality spectra and spectroscopic 
stellar parameters are available. The additional stars were chosen in a way to fill the parameter's range of 
applicability, that is, from F to K stars, from dwarfs to giants, and from metal-poor to metal-rich stars. 
Figure \ref{Fig4} shows the direct comparison of the derived spectroscopic parameters with the ones presented 
in Table \ref{tab1} for the benchmark stars, and with values derived in previous works (Sousa et al. 2008, 
2011a, 2011b) for the additional stars. The parameters derived with the new line-list and the comparison 
parameters are presented in Table \ref{tab2}. The errors were derived as in previous works (for details 
see \citet[][]{Sousa-2011a} and references within).

The temperatures and [Fe/H] show very consistent results within the error bars. For 
log g, the consistency is good in general, but there are specific cases that are not covered by the 
error bars. For these cases it is interesting, however, that the temperatures and [Fe/H] remain consistent and 
show little dependence on the derived surface gravity.

This procedure can be used to define a common line-list for different groups using an 
EW method. Something similar could also be used to define a good set of lines to derive homogeneous and 
accurate abundances for other elements if there are good benchmark abundances. The procedure can also be used or 
easily adapted for other surveys or even for the individual analysis of any given solar-type star.

\section{Summary}

The work presented can be essential for homogeneously determining spectroscopic parameters and chemical 
abundances for any large spectroscopic survey data using an EW method. 

We presented a technical overview of the standard spectroscopic method used by our team, which is based on 
the measurement of EWs of iron lines. In previous works our team has been using a differential approach, 
which cannot be adopted if strong constraints are imposed. To overcome this situation, we proposed here 
an alternative procedure that allows one to perform an optimal selection of the lines.

Using a small sample of stars that cover different ranges of stellar parameters, we showed that the use of the 
selected final line-list together with our ARES+MOOG method derives homogeneous and consistent
spectroscopic parameters.

To ensure a homogeneous derivation of spectroscopic stellar parameters, a final and 
fixed line-list can be obtained with this procedure to be used by other groups that use the EW method.

\begin{acknowledgements}
S.G.S, EDM, and V.Zh.A. acknowledge support from the Funda\c{c}\~ao para a Ci\^encia e Tecnologia (Portugal) in the form of 
grants SFRH/BPD/47611/2008, SFRH/BPD/76606/2011, SFRH/BPD/70574/2010, respectively . NCS thanks for the support by the European Research Council/European Community 
under the FP7 through a Starting Grant. We also acknowledge support from FCT and FSE/POPH in the form of grants 
reference PTDC/CTE-AST/098528/2008, PTDC/CTE-AST/098754/2008, and PTDC/CTE-AST/098604/2008. We are also greatful 
for the support of FP7 through the GREAT-ITN project FP7-PEOPLE-2010-ITN \& PITN-GA-2010-264895. The results presented 
here benefited from discussions in three Gaia-ESO workshops supported by the ESF (European Science Foundation) through 
the GREAT (Gaia Research for European Astronomy Training) initiative (Science meetings 3855, 4127 and 4415), HT acknowledges
the Ministerio de Economía y Competitividad (MINECO) under grants AYA2011-30147-C03-02* and BES-2009-012182, and the ESF and GREAT for a
n exchange grant 4158. HT and DM acknowledge The Universidad Complutense de Madrid (UCM) and The Comunidad de Madrid under PRICIT Project 
AstroMadrid (CAM S2009/ESP-1496). J.I.G.H. acknowledges financial support from the Spanish Ministry project MINECO AYA2011-29060, and
also from the Spanish Ministry of Economy and Competitiveness (MINECO) under the 2011 Severo Ochoa Program MINECO SEV-2011-0187. 
AJK acknowledges support by the Swedish National Space Board (SNSB).

\end{acknowledgements}

\bibliographystyle{bibtex/aa}
\bibliography{sousa_bibliography}

\end{document}